\documentclass[twocolumn,prl,superscriptaddress]{revtex4-1}

\usepackage{graphicx}
\usepackage{amsmath}
\usepackage{amsfonts}
\usepackage{amssymb}

\usepackage{textcomp}

\renewcommand{\t}[1]{\textrm{#1}}

\newcommand{\nn}{\nonumber\\}

\begin{document}
\title{Nonlinear cavity feeding and unconventional photon statistics in solid-state cavity QED 
revealed by many-level real-time path-integral calculations}

\author{M. Cygorek}
\affiliation{Theoretische Physik III, Universit{\"a}t Bayreuth, 95440 Bayreuth, Germany}

\author{A. M. Barth}
\affiliation{Theoretische Physik III, Universit{\"a}t Bayreuth, 95440 Bayreuth, Germany}

\author{F. Ungar}
\affiliation{Theoretische Physik III, Universit{\"a}t Bayreuth, 95440 Bayreuth, Germany}

\author{A. Vagov}
\affiliation{Theoretische Physik III, Universit{\"a}t Bayreuth, 95440 Bayreuth, Germany}

\author{V. M. Axt}
\affiliation{Theoretische Physik III, Universit{\"a}t Bayreuth, 95440 Bayreuth, Germany}

\begin{abstract}
The generation of photons in a microcavity coupled to 
a laser-driven quantum dot interacting with longitudinal acoustic (LA) phonons
is studied in the regime of simultaneously strong driving and strong 
dot-cavity coupling. The stationary cavity photon number 
is found to depend in a non-trivial way on the detuning between the laser 
and the exciton transition in the dot. 
In particular, the maximal efficiency of the cavity feeding is obtained 
for detunings corresponding to transition energies between
cavity-dressed states with excitation numbers larger than one. 
Phonons significantly enhance the cavity
feeding at large detunings. 
In the strong-driving, strong-coupling limit, the
photon statistics is highly non-Poissonian. While without phonons
a double-peaked structure in the photon distribution is predicted, 
phonons make the photon statistics thermal-like with very
high effective temperatures $\sim 10^5$ K, even for low phonon temperatures
$\sim 4$ K.
These results were obtained by numerical calculations
where the driving, the dot-cavity coupling and the dot-phonon 
interactions are taken into account without approximations.
This is achieved by a reformulation of an exact iterative path-integral scheme
which is applicable for a large class of quantum-dissipative systems and 
which in our case reduces the numerical demands by 15 orders of magnitude. 
\end{abstract}

\maketitle

Solid-state quantum dots (QDs) have attracted much attention in recent years
since they promise applications in photonic devices and  quantum
information technology, e.g., as qubits \cite{qubit_dot_Sham} and sources of single
\cite{single-photon3,single-photon1,single-photon2} or entangled
photons \cite{ncomm_dalacu2014, entangled-photon1, entangled-photon2}. 
Embedding QDs in microcavities increases the
light extraction efficiency via the Purcell effect \cite{single-photon3} 
and allows one to study cavity QED in solid-state systems 
\cite{DotInPhotonicCrystal,
hennessy07,reithmaier08,jaynes-cummings_ladder,Reitzenstein12,
Kasprzak13,
Hughes_master_equation_polaron_cavity,
Hughes_QDcavity_strongdriving_2011,Reitzenstein_QDcavity_strongdriving_2011}.
Cavities may be used as busses mediating a selective coupling of two qubits 
stored in two QDs within the same
cavity \cite{cQED_DAmico,cQED_Imamoglu_PRL,cQED_Imamoglu}.
Investigating cavity QED in a solid is a rich field, also
because the interaction with LA phonons
may have a profound impact on the physics \cite{Nahri,Kaer2010,Kaer},
e.g., enabling phonon-mediated inversion of the QD \cite{HughesCarmichael2013}  
or providing a non-resonant coupling between QD and cavity \cite{Ates2009}.

Recently, experiments \cite{cQED_Reitzenstein2017} on solid-state cavity
QED systems have advanced into the largely unexplored regime of
strong dot-cavity coupling combined with strong laser driving.
While strongly driven dots can be described in terms of laser-dressed
states
and the physics of strongly coupled dot-cavity systems is best discussed
in the basis of the cavity-dressed states, 
developing a physical intuition is more difficult
when driving and coupling are equally strong.
A further challenge is the coupling of the QD to phonons, in particular since
phonon influences on the QD dynamics are typically not well
described by Born-Markov rate equations \cite{McCutcheonPI}. Thus, 
more sophisticated approaches are required, like the
correlation expansion \cite{Knorr2003,PI_phonon-assisted,Doris2017} 
or polaron master equations \cite{NazirReview,Kaer,Hughes_master_equation_polaron_cavity,Pawel2015,Iles-Smith2016}.

\begin{figure}
\includegraphics[width=0.35\textwidth]{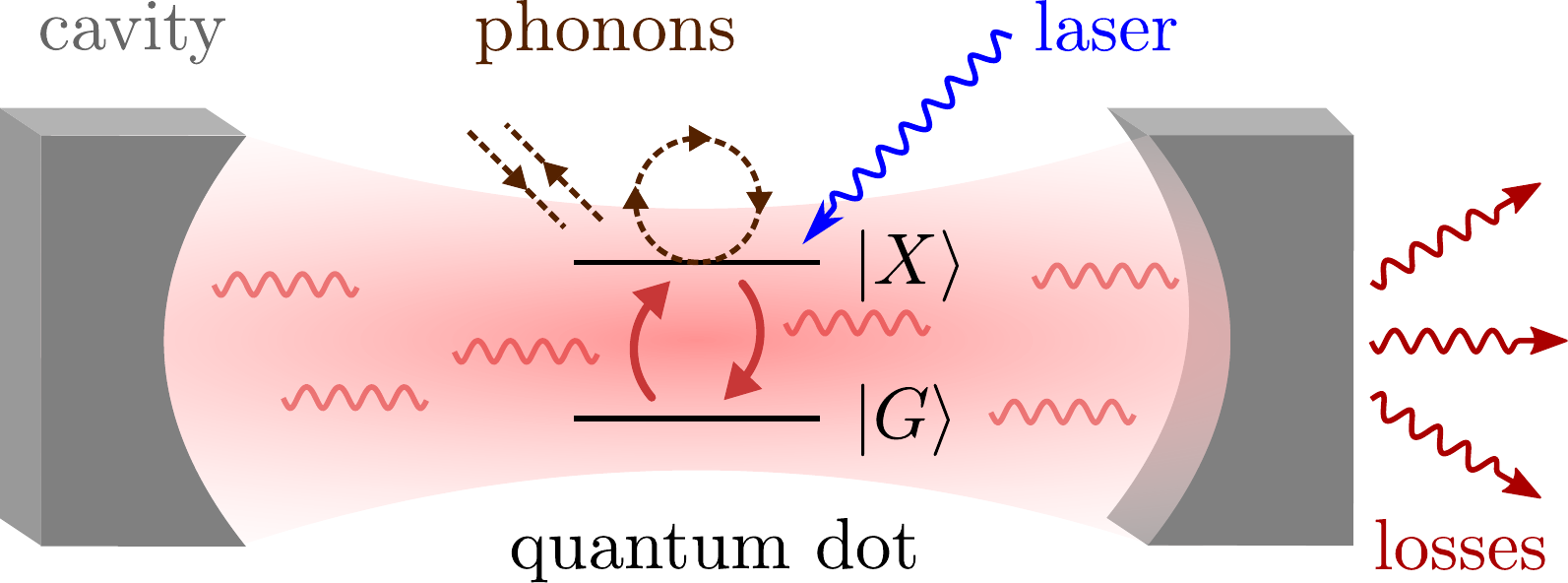}
\caption{\label{fig:sketch}
Laser-driven quantum dot inside a microcavity and coupled to 
longitudinal acoustic phonons.
}
\end{figure}

For possible applications of strongly coupled dot-cavity systems, e.g., as building blocks of 
photonic devices, it is of fundamental interest how efficiently photons  
can be generated in the microcavity by driving the QD with an external 
laser field.
In this letter, we address  this question for cases where QD-cavity and QD-laser couplings are simultaneously
strong while 
also accounting for the QD-phonon interaction.
Intuition suggests that the cavity feeding is enhanced when
the laser is tuned to the resonances in the linear absorption of the
dot-cavity system, i.e., when the dot-laser detuning matches the vacuum-Rabi-split 
peaks corresponding to the transitions between the ground state and the 
first excited cavity-dressed states.
We demonstrate that this expectation is only confirmed
at low driving strength where the average cavity photon number is much smaller than one due to
the photon blockade \cite{PhotonBlockade2008}.
At strong driving, we find the maximum feeding efficiency at 
much smaller detunings, 
while there is  no enhancement at the resonances of the linear absorption.

Phonons are a major reason why solid-state cavity QED differs from atomic cavity QED.
In most cases phonons limit the performance of device relevant processes by introducing
decoherence. However, in this letter we demonstrate that,
already for small detunings compared to the vacuum Rabi splitting, 
phonons lead to a strongly enhanced generation of cavity photons that becomes almost
independent of the detuning and all traces of the photon blockade are
eliminated.

While in weak-coupling or weak-driving situations only cavity
states with one or two photons can be significantly occupied, 
strong-driving+strong-coupling conditions lead to the excitation
of states with larger photon numbers. This makes the shape of the 
photon distribution a meaningful target for investigations.
In contrast to the case of classical (direct) driving of a cavity,
which leads to a Poissonian photon distribution \cite{Glauber}, we find
strongly non-Poissonian distributions when the cavity is driven 
indirectly via the dot.
Here, phonons not only lead to drastic quantitative effects but change the 
photon statistics qualitatively by transforming the
photon distribution into a nearly thermal one.

\begin{figure*}[ht]
\includegraphics[width=\textwidth]{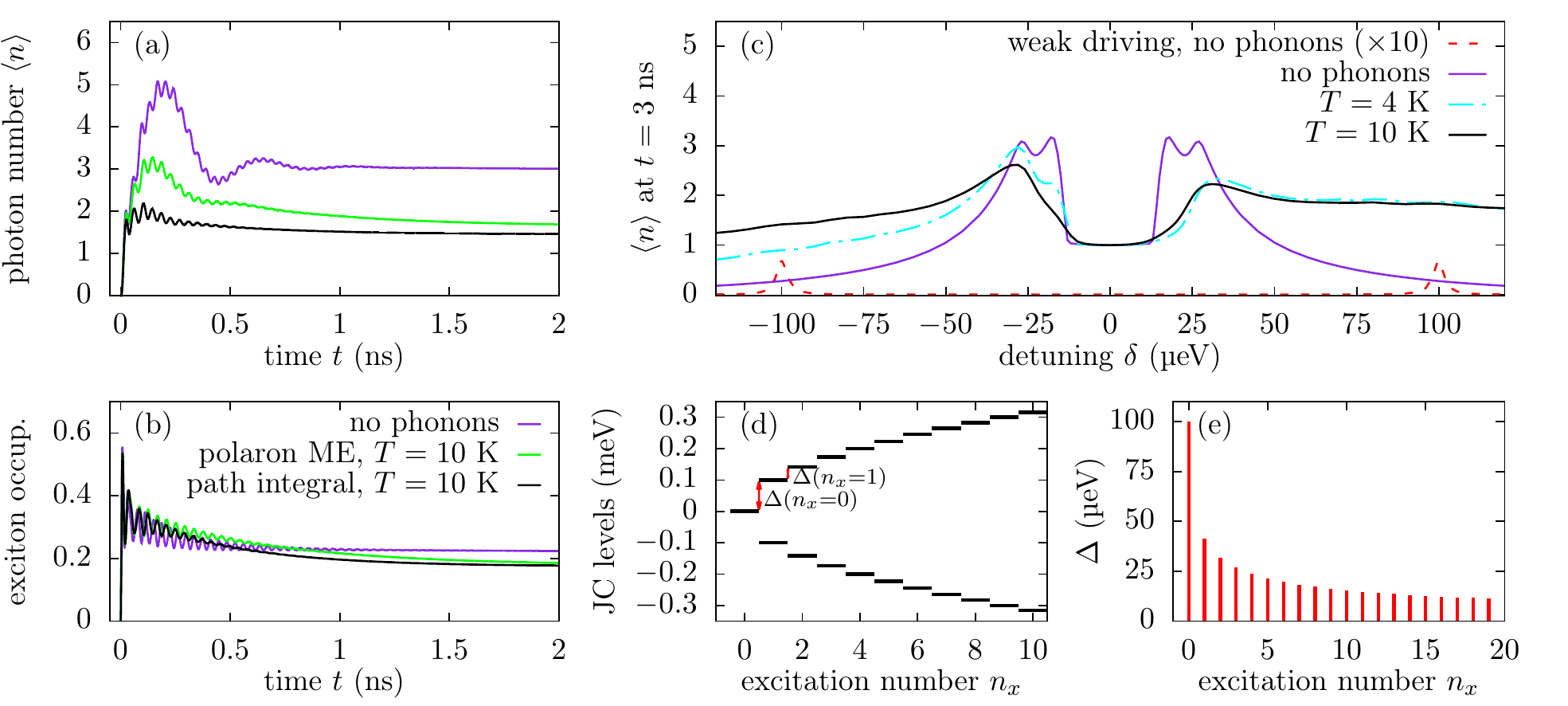}
\caption{\label{fig:mult1}
Time evolution of the average photon number (a) and of the exciton occupation (b)
for a detuning $\delta=20$ \textmu eV at $T=10$ K 
calculated using the path-integral theory (black), 
Markovian polaron master equations (green), and without phonons (violet).
(c): Average photon number $\langle n\rangle$ in the cavity at  $t=3$ ns
as a function of the detuning $\delta$.
(d): Energy of the Jaynes-Cummings levels in the rotating frame.
(e): Transition energies $\Delta(n_x)$ between neighboring Jaynes-Cummings
levels [as indicated in (d)].}
\end{figure*}

We study a QD-cavity system interacting with LA phonons considering a single
ground-to-exciton transition of the QD coupled to a single cavity mode 
and driven by an external laser field as sketched in Fig.~\ref{fig:sketch}. 
%
%
The Hamiltonian for the dot-cavity system 
in the interaction picture with respect to the laser energy
$\hbar\omega_L$ is given by
\begin{align}
H_N=&\hbar\Delta\omega_{XL} |X\rangle\langle X| 
- \hbar f(|G\rangle\langle X|+|X\rangle\langle G|)
\nn&
+ \hbar\Delta\omega_{cL} \hat{a}^\dagger \hat{a} 
+ \hbar g(\hat{a}^\dagger |G\rangle\langle X|+\hat{a} |X\rangle\langle G|)
,
\label{eq:HN}
\end{align}
where $|G\rangle$ $(|X\rangle)$ denotes the ground (excited) state of the QD 
and $\hat{a}^\dagger$ ($\hat a$) is the creation (annihilation) operator
of a cavity photon.
$\hbar\Delta\omega_{XL}=\hbar(\omega_{X}-\omega_{L})$ 
and $\hbar\omega_X$ is the exciton transition energy
while $\hbar\Delta\omega_{cL}=\hbar\omega_c-\hbar\omega_L$ 
is the detuning of the cavity with respect to the external driving.
$g$ is the strength of the dot-cavity coupling
and $f$ denotes the strength of the external cw laser driving.

LA phonons are included by the Hamiltonian \cite{besombes:01,pawel04}
\begin{align}
\hat{H}_{ph}=\hbar\sum_{\mathbf q}\omega_{\mathbf q}\hat{b}^\dagger_{\mathbf q} \hat{b}_{\mathbf q}
+\hbar\sum_{\mathbf q}\big(\gamma_{\mathbf q}^X \hat{b}^\dagger_{\mathbf q}+\gamma_{\mathbf q}^{X*}\hat{b}_{\mathbf q}\big)
|X\rangle\langle X|,
\label{eq:Hosc}
\end{align}
where $\hat{b}^\dagger_{\mathbf q}$ and $\hat{b}_{\mathbf q}$ are creation and 
annihilation operators for
phonons with energy $\hbar\omega_{\mathbf q}$. $\gamma_{\mathbf q}^X$ are 
the exciton-phonon-coupling matrix elements given explicitly in the
supplement.
Finally, cavity losses are taken into account via the Lindblad term
$\mathcal{L}_{\text{loss}}[\hat{\rho}]=
\kappa\big[\hat{a} \hat\rho\hat{a}^\dagger
-\frac 12\{\hat\rho, \hat{a}^\dagger \hat{a}\}_+\big],$
with loss rate $\kappa$.

The number of cavity photons generated by driving via the dot is limited
by the losses. However, when the driving ($f\gg\kappa$) and the QD-cavity
coupling ($g\gg\kappa$) are strong, describing  the dynamics requires
accounting for a large number of states of the QD-cavity system. 
Here we consider all dot-cavity product states with up to $n_x\le 20$ 
excitations, where the excitation number $n_x$ is the photon number $n$ 
plus the exciton occupation. Since there is one pair of states 
$(|G,n_x\rangle,|X,n_x-1\rangle)$ per excitation $n_x>0$ and only one state
$|G,0\rangle$ with $n_{x}=0$, this amounts to a system with $N=41$ levels.

It is important to note that strongly-driven, strongly-coupled solid-state 
cavity QED systems are simultaneously highly non-linear 
with respect to the driving, the dot-cavity coupling, and the dot-phonon 
coupling, so that there is no obvious small parameter in the system. 
As a consequence, it is {\em a priory} unclear whether conclusive results 
can be obtained from established approximate methods, such as 
master equation or correlation expansion approaches.
In principle, the time evolution of the
reduced density matrix of an $N$-level system coupled to a continuum of phonons
can be calculated using numerically exact path integral (PI) methods 
\cite{Makri_Theory,Makri_Numerics,PI_realtime2011,Nahri,PI_nonHamil2016}.
Unfortunately, the numerical effort required in such calculations
rises exponentially with the number of levels $N$, so that 
complete PI simulations have been performed only for rather small systems 
\cite{LeggetRevModPhys,Makri_3lvl,Thorwart_Landau-Zener,Nahri,
PI_undressing,PI_phonon-assisted_biexc_prep,polarization}. In particular,
when the phonon-induced memory time is as long as a few picoseconds
(which is typical for LA phonons coupled to a  QD) 
numerically complete simulations are usually restricted to $N\le4$,
while larger systems may be treated 
by heuristically discarding a large number of paths with numerically small 
contributions \cite{on-the-fly-filtering,Makri_Blip,Multilevel}.
For solid-state cavity QED systems with $N\sim 40$ levels,
complete PI simulations have, so far, been performed only for lossless 
cavities without external driving \cite{PI_Cavity_Soergel}, where the quantum
dynamics of the system mixes only pairs of states with fixed excitation 
numbers $n_x$.
However, external driving or cavity losses introduce transitions between states 
with different excitation numbers, so that the fully coupled system has to be 
considered, which increases the computation time dramatically.

We have overcome this problem by reformulating a standard  
iterative algorithm \cite{Makri_Theory} to perform the sum over the paths, which in our case reduces the number
of entities that have to be iterated by more than 15 orders of magnitude without introducing
approximations. 
Full details of the reformulated path-integral method and
its derivation are given in the supplement.
Here, we only note that the reformulation can be applied generally to any $N$-level system
coupled to an oscillator continuum, provided the $N$ states can be subdivided into $N_{g}$ groups
with identical oscillator couplings within each group. 
In our case, the interaction with phonons
does not distinguish between states that differ only in the number of cavity photons.
Thus, we have $N_{g}=2$ were one group comprises the states $|G,n_x\rangle$ while a second group
contains the states $|X,n_x-1\rangle$. 

It is worthwhile to note that apart from the specific case treated in this letter there is
a wealth of other systems of topical interest where the new algorithm can be applied.
An example of such systems are QDs with embedded magnetic dopants, e.g.,
Mn ions, which are highly attractive for spintronic applications 
\cite{besombes:04,kudelski:07,MnSpin}. Here, the phonon coupling does not 
distinguish between different spin configurations of the dot-dopant system.
Another possible application is the description of phonon effects
on the biexciton cascade in a QD, proposed as a source of
entangled photon pairs \cite{ncomm_dalacu2014, entangled-photon1, 
entangled-photon2}. As in our system, the phonon coupling does not distinguish between
states differing only by photon numbers such that for the biexciton cascade
$N_g=3$ groups have to be considered  (ground state, excitons, and biexciton). 
Cavity QED has also been studied in systems where superconducting
charge qubits are strongly coupled to a microwave cavity \cite{cQED_supercond}.
There, the dephasing is determined by charge fluctuations that can 
be represented by a bath of harmonic oscillators \cite{Makhlin_RevModPhys},
so that our PI method can be applied to these systems
as well.

For the present study of cavity feeding in
QD-based solid-state cavity QED systems, 
we assume a cavity 
in resonance with the polaron-shifted
QD-exciton transition and a laser coupled to the QD transition
detuned by an energy $\delta$ 
from the cavity mode.
Before the laser is switched on at time $t=0$ the 
electronic system is in the ground state $|G,0\rangle$ while the 
phonons are initially in thermal equilibrium at temperature $T$.
The dot-cavity coupling and the driving are chosen to be
of equal strength $\hbar g=\hbar f=0.1$ meV, if not stated otherwise. 
The cavity loss rate 
is taken to be $\kappa=0.01$ ps$^{-1}$, which corresponds to a quality factor
 $Q\approx 10^5$. For the phonon environment and dot-phonon coupling
we assume parameters of a self-assembled \mbox{InGaAs} QD with 
radius $a_e=3$ nm embedded in a GaAs matrix (cf. supplement).

\begin{figure*}[ht]
\includegraphics[height=4.4cm]{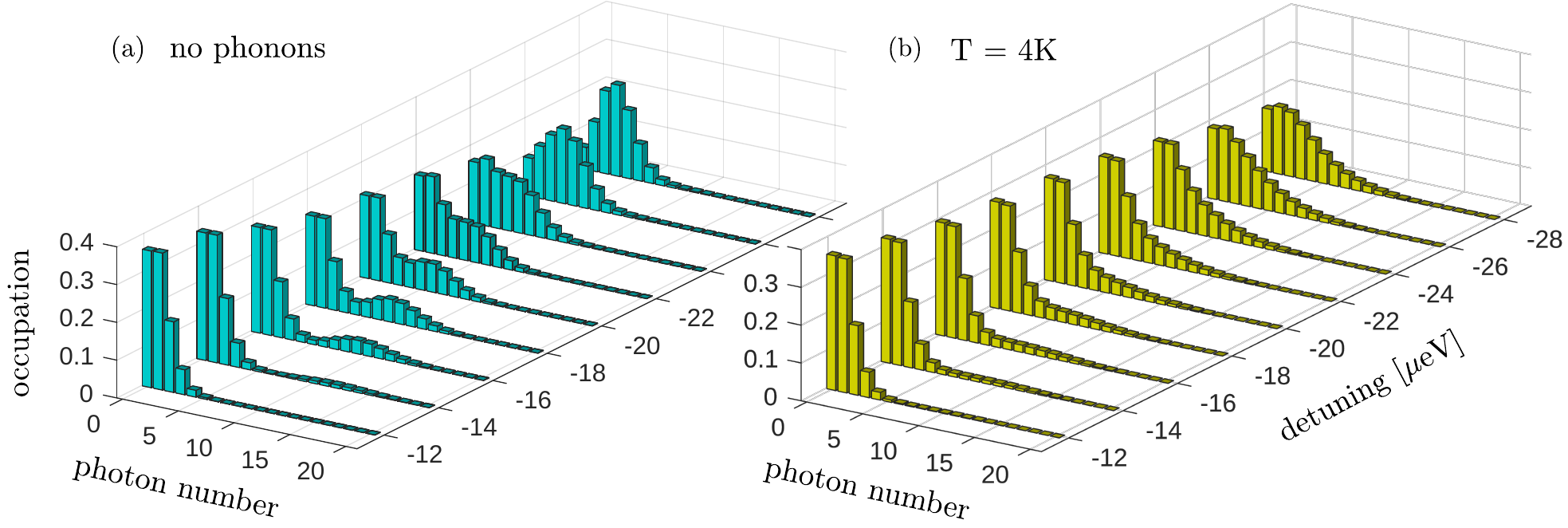}
\hfill
\includegraphics[height=4.4cm]{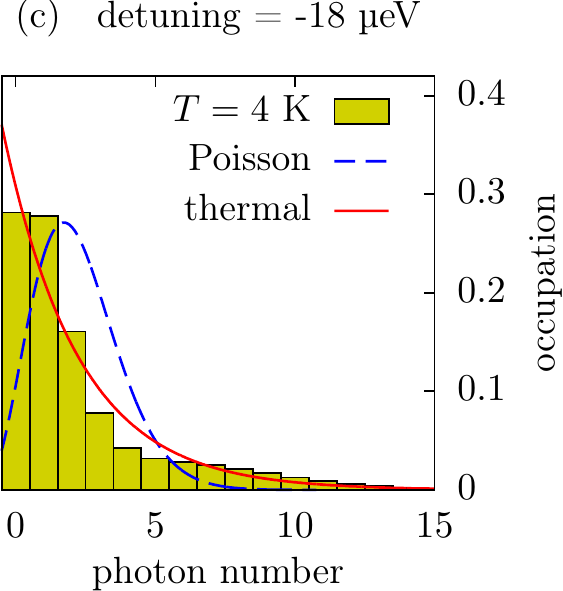}
\caption{\label{fig:histo}
Cavity photon distribution at $t=3$ ns for different detunings $\delta$ 
and $\hbar g=\hbar f=0.1$ meV without 
dot-phonon interaction (a) and with phonons at temperature $T=4$ K (b).
(c): Photon distribution at detuning $\delta=-18$ \textmu eV with phonons 
at $T=4$ K compared with Poissonian and thermal distributions.
}
\end{figure*}

Figures~\ref{fig:mult1}(a) and (b) depict the time evolution of the 
average cavity photon number $\langle n\rangle$ and the exciton occupation
for a detuning $\delta=20$ \textmu eV. 
Both quantities show an oscillatory transient behavior in the first $\sim$ 1 ns
and eventually reach stationary values.
We find that phonons have little impact on the exciton occupations
but can change the average photon number significantly.
For comparison, we present results obtained using polaron master equations 
(PMEs, cf. supplement), a well established method
for the treatment of the dot-phonon interaction \cite{Iles-Smith2016,
NazirReview,McCutcheonNazir_beyondWC,Hughes_master_equation_polaron_cavity}.
These results demonstrate that, while exciton occupations are reasonably well described by
the PMEs, photon related quantities are much more
affected by the approximations used to derive the PMEs.
In particular, the PMEs predict significantly larger 
(e.g., more than $60\%$ at $t=150$ ps)
photon numbers than the PI method, 
revealing that the PMEs are insufficient for an accurate description
of the situation studied here.
We note that this inaccuracy of the PMEs can be explained by the fact that
the dot-cavity system possesses a number of different energy gaps on scales 
$\sim$ 10--1000 \textmu eV [cf. Fig. \ref{fig:mult1}(d)], which can be 
bridged by phonon-assisted transitions. These are, however, only
strictly treated up to second order in the dot-phonon interaction
within the PME approach.

Phonon effects on the cavity feeding are illustrated
in Fig.~\ref{fig:mult1}(c) showing the average photon number 
at long times ($t=3$ ns)
as a function of $\delta$ with and without dot-phonon interaction.
In the weak-driving limit ($\hbar f=1$ \textmu eV, red line), 
the driving of the cavity via the dot is
most effective if one excites 
at the resonances of the linear absorption, 
manifested in the appearance of two peaks 
at $\delta=\pm\hbar g=\pm 100$ \textmu eV
separated by the vacuum Rabi splitting. 
Surprisingly, for strong driving the maximum 
number of photons in the cavity is obtained neither at
$\delta=\pm\hbar g$ nor when the driving is in resonance with the 
ground-to-exciton transition ($\delta=0$, i.e., where the linear absorption
has a resonance for vanishing dot-cavity coupling), but in a region with
small detunings of $\sim$ 20--40 \textmu eV.
This is explained by the fact that with strong driving the
cavity is partially filled with photons and thus transitions between
states with higher photon numbers become important. When the dot and the 
cavity are in resonance, the energy eigenvalues of the cavity-dressed states 
are
$E_{n_x}= \pm \sqrt{n_x}\hbar g$ [cf. Fig~\ref{fig:mult1}(d)] and
transitions between neighboring states can occur at energies
$\Delta(n_x)=(\sqrt{n_x+1}-\sqrt{n_x})\hbar g$ depicted in
Fig.~\ref{fig:mult1}(e).
For the chosen parameters, the efficiency of the cavity feeding 
increases drastically when $\delta$ becomes similar to $\Delta(n_x)$
for $n_x\approx 5$, which corresponds to the typical photon number 
of states that are significantly occupied.
When the driving strength increases further and states with higher photon 
numbers are occupied, the maxima in Fig~\ref{fig:mult1}(c) 
will eventually shift to zero and merge into the
central peak of a Mollow triplet \cite{Mollow}.

Fig.~\ref{fig:mult1}(c) also reveals an asymmetry of the average photon number 
with respect to a sign change of the detuning.
This asymmetry diminishes at higher temperatures indicating that it originates
from the asymmetry between phonon absorption and emission. 
Compared with phonon-free calculations,
the interaction with phonons results in a reduced efficiency at the maxima
 because it suppresses the coherent driving. 
However, at larger detunings, phonon-assisted processes 
facilitate otherwise prohibited transitions between off-resonant states, which 
enables a much more efficient generation of cavity photons.
Interestingly, the phonon-induced feeding efficiency 
in this regime is almost independent of the detuning in a wide parameter 
range.

More detailed information than the mean number of photons is provided
by examining how the photons are distributed across different photonic states.
It is well known that direct (classical) driving of a cavity leads to a
Poissonian statistics \cite{Glauber}.
Here, however, we find that driving the cavity via the dot creates 
highly non-Poissonian distributions as depicted in 
Fig.~\ref{fig:histo}.
Figure~\ref{fig:histo}(a) illustrates the case where
the dot-phonon interaction is neglected. There, large detunings lead to 
a shift of the peak in the photon distribution to higher photon numbers 
with a rapidly decaying tail, as expected for a Poisson distribution.
For values of the detuning at which
the resonance condition with neighboring cavity-dressed energy
eigenstates is met, a double-peak structure appears in the photon statistics.
This two-peak structure is a rather unconventional feature, 
which is possible only due to the resonance in the
non-linear driving regime.

When the dot-phonon interaction is accounted for [cf.~Fig.~\ref{fig:histo}(b)],
significant changes in the photon distributions are observed. Most prominently,
the double-peak structure predicted by the phonon-free calculation disappears.
Furthermore, for large photon numbers, the photon distribution has a maximum 
closer to $n=0$ and possesses a significantly 
longer tail than expected for a Poissonian distribution.
In Fig.~\ref{fig:histo}(c), we compare the photon statistics 
for $\delta=-18$ \textmu eV, 
where the maximal efficiency of cavity feeding was reached, 
with a Poissonian as well as with a thermal distribution
[$P_\t{th}(n)=(1-e^{-\epsilon}) e^{-\epsilon n}$]  
with the same average photon number $\langle n\rangle\approx 2.25$ as 
obtained in the numerical simulations for $T=4$ K.
It can be seen that the obtained photon statistics in the presence of phonons 
is much closer to the thermal distribution than to the Poissonian. 
For a cavity mode with energy $\hbar\omega_c\approx 1.5$ eV, 
one can extract an effective temperature 
$T=\hbar\omega_c/(k_B \epsilon)\approx 47\,000$ K. 
It is worth noting that, although the qualitative change of the photon 
statistics toward a thermal distribution is caused by the 
dot-phonon interaction, the value of the photon temperature is four orders of 
magnitude larger than the phonon temperature and is rather determined
by the average photon number, which can be tuned by changing
the driving strength.

To summarize, we have investigated the generation of photons in a QD-cavity
system coupled to LA phonons in the regime of strong driving and 
simultaneously strong coupling.
Conclusive results have been obtained 
using a novel variant of the 
numerically exact real-time path-integral approach, 
which for a large class of systems of topical interest 
speeds up the numerics by many orders of magnitude. 
Our simulations  show that, when
a dot is in resonance with a microcavity, the feeding efficiency 
depends non-trivially on the dot-laser detuning. 
In the strong-driving limit, the maximal feeding is observed
for much smaller detunings than expected from the linear absorption. 
The dot-phonon interaction is found to suppress the feeding efficiency 
at resonances. However, for larger detunings, it
opens up the possibility of a highly efficient phonon-mediated feeding
which is robust against variations of the detuning.
Furthermore, the dot-phonon interaction modifies the photon statistics 
qualitatively, so that a double peaked distribution for the  phonon-free case is
transformed to a nearly thermal occupation of photon states 
with a huge effective temperature. 

\acknowledgements
We gratefully acknowledge the financial support from 
Deutsche Forschungsgemeinschaft via the Project No. AX 17/7-1.

\bibliography{PIbib}
\end{document}